\documentclass[10pt]{article}
\usepackage[latin1]{inputenc}
\usepackage{amsmath}
\usepackage{amsfonts}
\usepackage{amssymb}
\usepackage{mathrsfs}

\usepackage[authoryear]{natbib}
\usepackage[colorlinks=true, citecolor=black, linkcolor=black, urlcolor=blue]{hyperref}

\usepackage{blindtext}

\usepackage{MnSymbol}

\usepackage[cal=boondox,scr=boondoxo]{mathalfa}

\usepackage{float}
\usepackage{subfig}
\usepackage{fancybox,graphicx}
\usepackage{subfig}
\usepackage{caption}
\usepackage{color}
\usepackage{authblk}

\usepackage{accents}
\usepackage[titletoc,title]{appendix}
\usepackage{cite}

\usepackage{mathtools}

\usepackage[top=2in, bottom=1.5in, left=1in, right=1in]{geometry}

\usepackage{stackengine}

\title{Intelligent Product 3.0: Decentralised AI Agents and Web3 Intelligence Standards}
 
\author[1,2]{Alex C.Y. Wong }
\author[1]{D. McFarlane}
\author[2]{C. Ellarby}
\author[2]{M. Lee}
\author[2]{M. Kuok}

\affil[1]{University of Cambridge, Cambridge, UK}
\affil[2]{RedBite Solutions Ltd, Cambridge, UK\\\texttt{www.redbite.com}}

\begin{document}
\maketitle

\begin{abstract}
Twenty-five years ago, the specification of the Intelligent Product was established, envisaging real-time connectivity that not only enables products to gather accurate data about themselves but also allows them to assess and influence their own destiny. Early work by the Auto-ID project focused on creating a single, open-standard repository for storing and retrieving product information, laying a foundation for scalable connectivity. A decade later, the approach was revisited in light of low-cost RFID systems that promised a low-cost link between physical goods and networked information environments. Since then, advances in blockchain, Web3, and artificial intelligence have introduced unprecedented levels of resilience, consensus, and autonomy. By leveraging decentralised identity, blockchain-based product information and history, and intelligent AI-to-AI collaboration, this paper examines these developments and outlines a new specification for the Intelligent Product 3.0, illustrating how decentralised and AI-driven capabilities facilitate seamless interaction between physical AI and everyday products.
\end{abstract}

\begin{quote}
\noindent \text{Keywords:} Intelligent Products, Auto-ID, Web3, Decentralisation, AI, Blockchain, Physical AI, Intelligent Agents, Multi-Agent System (MAS), Digital Product Passport, IoT, RFID, DePIN
\end{quote}

\section{Introduction}

The ``Intelligent Product'' was first introduced as a way to embed intelligence within everyday objects, enabling them to assess and influence their own destiny \citep{Wong2002Intelligent}. The concept built on the technologies and infrastructure being developed at the Auto-ID Center \citep{Sarma2000NetworkedWorld}, notably the Electronic Product Code (EPC) for Radio Frequency Identification (RFID), along with related standards for storing and communicating product data. However, this predated blockchain, while the Internet of Things (IoT), a term also coined at the Auto-ID Center by Kevin Ashton \citep{ashton2009iot}, and the Internet itself were still in their infancy as communication platforms. Embedded AI, primarily implemented through software agents, remained largely a research tool at the time. As a result, truly autonomous and fully intelligent products were not attainable until recent innovations in blockchain, Web3, and artificial intelligence. This paper revisits the original vision and specification of the Intelligent Product, charts its refinement over the years, and demonstrates how these emerging capabilities have paved the way for Intelligent Product 3.0.

\section{The Intelligent Product: An Overview}

This section provides a brief overview of the original Intelligent Product definition and examines how it has evolved over time. At the end of this section, we identify the prevailing challenges and limitations within the current framework, thereby establishing the context for further technological advancement of the Intelligent Product.

\subsection{The Intelligent Product Definition}

The original definition of the Intelligent Product as defined in \citet{Wong2002Intelligent} and subsequently in \citet{McFarlane2003AutoID} is as follows:

\paragraph{Definition 1: Intelligent Product \citep{Wong2002Intelligent}}
\textit{An Intelligent Product is a product (or part or order) that has part or all of the following five characteristics:}
\begin{enumerate}
    \item Possesses a unique identity
    \item Is capable of communicating effectively with its environment
    \item Can retain or store data about itself
    \item Deploys a language to display its features, production requirements etc.
    \item Is capable of participating in or making decisions relevant to its own destiny
\end{enumerate}

The definition above does not prescribe specific technology implementations, rather it offers a conceptual foundation that can be realised through various technical means. A particular implementation discussed in the paper was based on the standards that were being developed at the Auto-ID Center at that time. The Auto-ID Center is a global initiative consisting of Massachusetts Institute of Technology (MIT) and University of Cambridge, along with other university labs subsequently.

The mission of the Auto-ID Center was to merge physical material flows (``atoms'') with networked information (``bits'') by embedding a unique identifier on a low-cost RFID tag \citep{Sarma2001FiveCentTag} attached to each product. This identifier, referred to as the Electronic Product Code (EPC) \citep{Brock2001EPC}, points to a network-based repository and a decision-making software agent. Working in tandem with the EPC, a proposed standard called Product Markup Language (PML) uses an XML-based format to provide a rich, dynamic data model for describing physical objects online. An Object Naming Service (ONS), partly based on the Internet's Domain Name System (DNS), then directs computer systems to the corresponding data repository for any item bearing an EPC code. Designed to handle information for trillions of tagged objects, this ``lightning-fast post office'' is anticipated to exceed the DNS in scale \citep{McFarlane2002IPMC}. We note that, although some of these standards, namely PML and ONS, did not achieve widespread adoption, their underlying principles remain foundational across a range of contemporary implementations.

\begin{figure}[H]
    \centering
    \includegraphics[width=0.5\textwidth]{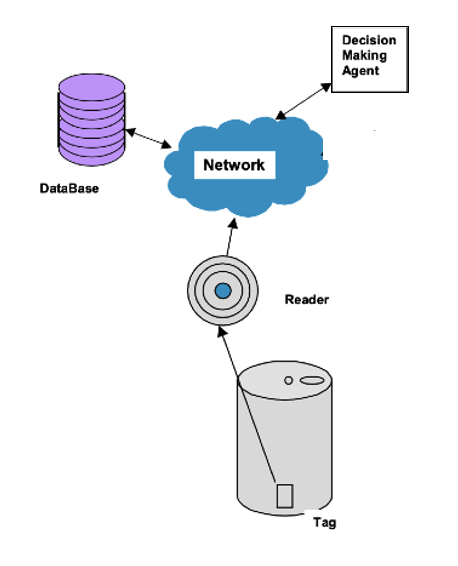}
    \caption{Intelligent jar of spaghetti sauce}
    \label{fig:sauce}
\end{figure}

A simplistic example of an ``intelligent'' spaghetti sauce jar  \citep{Wong2002Intelligent} was provided to illustrate the notion (see Figure~\ref{fig:sauce}).

The intelligent jar of spaghetti sauce, identifiable via a unique tag and capable of participating in a decision-making process through the network and database, exemplifies the integration of product intelligence. There were also two different types of product intelligence defined in the paper, as follows:

\paragraph{Definition 2: Levels of Product Intelligence \citep{Wong2002Intelligent}}
\begin{itemize}
    \item \textbf{Level 1 Product Intelligence:} which allows a product to communicate its status (form, composition, location, key features), i.e.\ it is information-oriented.
    \item \textbf{Level 2 Product Intelligence:} which allows a product to assess and influence its function (e.g.\ self-distributing inventory and self-manufacturing inventory) in addition to communicating its status, i.e.\ it is decision-oriented.
\end{itemize}

Level 1 Product Intelligence has been challenging to implement, yet it has seen the most activity, evolving from the now-deprecated Physical Markup Language (PML), an attribute-based XML framework, to Electronic Product Code Information Services (EPCIS), an event-driven standard for supply chain visibility, and more recently, to Digital Product Passport (DPP), a digital framework for capturing a product's lifecycle information.

Fundamental to Level 2 Product Intelligence is the ability of a product to make or influence decisions relevant to its own destiny, transforming a passive object into an autonomous entity capable of dynamic decision making and collaboration across networks \citep{Karkkainen2003Intelligent}. However, Level 2 intelligence has seen limited development to date. 

From the outset, \citet{Wong2002Intelligent} defined the use of software agents as the mechanism to achieve these capabilities, describing a software agent as:

\begin{quotation}
``A distinct software process, which can reason independently, and can react to change induced upon it by other agents and its environment, and is able to cooperate with other agents.''
\end{quotation}

Notably, no established standards or implementations for software agents or similar intelligent systems existed within the initial Auto-ID Center and subsequent GS1 framework. It is only with recent advancements in Agentic AI and multi-agent systems for large language models (LLMs) that this concept has gained renewed relevance. The implications of these advancements and their potential role in shaping Level 2 Product Intelligence will be further explored in Section~\ref{sec:3}.

\subsection{Deconstructing the Intelligent Product's Architecture}

The development of intelligent products represents a fundamental shift in how physical objects interact with digital systems. Traditionally, products were identified using static identifiers such as barcodes and RFID tags, which provided basic tracking capabilities but lacked contextual awareness and decision-making functions. In this section, we discuss the key enablers of an Intelligent Product, as defined in Definition 1, which enables the transition from static identification to autonomous intelligence.

\subsubsection{Globally Unique Identity: The Core Foundation of Intelligent Products}

A core requirement for intelligent products is the ability for these products to be uniquely identified within digital and physical ecosystems. Without a globally standardised identity, products cannot be tracked effectively, authenticated, or integrated into automated systems. The development of Auto-ID technologies has been instrumental in enabling machine-readable identification, providing the foundation for automated supply chains, asset tracking, and intelligent decision-making \citep{Framling2007UniqueID}.

The Electronic Product Code (EPC), introduced by the Auto-ID Center, was one of the earliest scalable frameworks designed to standardise unique identification across global supply chains \citep{Sarma2000NetworkedWorld}. EPC builds upon existing systems such as the Global Trade Item Number (GTIN) which is used extensively in retail and logistics. Several international standards further refine the rules governing product identity, including ISO/IEC 15459 \citep{ISO15459}, which defines globally unique identification codes, and GS1 EPC Information Services (EPCIS) \citep{GS1EPCIS2016}, which facilitates event-driven data exchange between businesses. Additionally, ISO/IEC 18000 \citep{ISO18000} specifies standards for RFID-based object identification, ensuring interoperability across different RFID systems. These frameworks provide the basis for machine-readable and networked identification, allowing physical products to be seamlessly integrated into digital infrastructures.

\subsubsection{Product Data: Capturing, Representing, Structuring, and Exchanging Information}

Beyond unique identification, intelligent products require a structured approach to capturing, storing, and exchanging data about their status and environment. The transition from passive identification to context-aware products necessitates the adoption of standardised data models, real-time sensing technologies, and interoperable communication protocols. To support this, intelligent products must address multiple aspects of data representation and management, including data formats, schemas, storage solutions, query languages, and mechanisms for secure data exchange \citep{Lopez2011SmartObjects}.

To ensure consistency in data representation, several international standards define how product data should be structured and shared across different platforms. ISO 10303 STEP \citep{ISO10303} provides a standardised format for exchanging 3D models and digital twins in manufacturing, ensuring compatibility between different CAD and product lifecycle management (PLM) systems. Similarly, GS1 EPCIS enables event-based data tracking, allowing businesses to capture product movements, temperature variations, and handling conditions in real time. In industrial applications, ISO 23386 \citep{ISO23386} and ISO 23387 \citep{ISO23387} establish a framework for digital product data exchange, supporting automation in Industry 4.0 environments. These standards facilitate the harmonisation of product data, allowing intelligent systems to interpret and respond to environmental changes more effectively.

Intelligent products also require real-time data acquisition technologies to enhance situational awareness. IoT sensors ISO/IEC 30141 \citep{ISO30141} enable continuous monitoring of temperature, pressure, humidity, and vibration, providing critical data for predictive maintenance and asset tracking. Additionally, GPS-based geolocation tracking plays a crucial role in smart logistics, fleet management, and mobility services, ensuring that products and assets can be located dynamically. Embedded computing standards such as IEEE 21451 \citep{IEEE21451} enable smart sensors to process data locally, facilitating edge computing architectures where intelligent products can perform real-time analysis without relying solely on cloud infrastructure.

Interoperability remains a key challenge in intelligent product ecosystems, necessitating standardised communication protocols for seamless data exchange. MQTT ISO/IEC 20922 \citep{ISO20922} serves as a lightweight messaging protocol widely used in IoT applications, enabling efficient data transfer between sensors, devices, and cloud systems. OPC UA IEC 62541 \citep{IEC62541}
 provides a secure machine-to-machine communication framework, particularly in industrial automation settings, ensuring that products can interact across different manufacturing systems. Furthermore, ISO/IEC 20248 \citep{ISO20248} introduces digital signatures for product data, enhancing security and trust in digital transactions by ensuring that product-related information remains tamper-proof.

\subsubsection{Product Intelligence: Enabling Autonomous Decision-Making}

The highest level of intelligent product development is characterised by autonomous reasoning, real-time responsiveness, and multi-agent coordination. Unlike traditional automated systems, which rely on predefined rules and centralised control, Intelligent Products are expected to independently assess environmental conditions and adapt their behaviour accordingly. This is achieved through agent-based architectures, multi-agent collaboration, and distributed computing models.

An agent-based system is a computational framework in which autonomous software entities (agents) interact dynamically with their environment and with other agents to achieve specific objectives \citep{Wong2002Intelligent}. Beyond individual agents, multi-agent systems (MAS) provide a structured approach for distributed, intelligent decision-making \citep{Jennings1998AgentsRoadmap}. MAS architectures enable multiple autonomous agents to coordinate their actions, facilitating decentralised problem-solving in environments such as manufacturing control, logistics, and industrial automation \citep{McFarlane2003AutoID}. This approach is particularly valuable in large-scale smart supply chain networks, where multiple intelligent products must interact dynamically to optimise inventory management, transportation, and predictive maintenance.

The implementation of intelligent products can follow two primary models: localised intelligence or network-based intelligence. In the localised approach, intelligence is embedded directly into the product, allowing it to operate autonomously without reliance on external systems. This is particularly useful for applications requiring low-latency decision-making, such as autonomous robots, industrial automation, and real-time monitoring systems. Alternatively, in the network-based approach, intelligence is distributed across cloud platforms and edge computing frameworks, enabling products to access real-time data, analytics, and decision-support models over a connected infrastructure \citep{Meyer2009Survey}. The choice between these models depends on the specific performance, connectivity, and scalability requirements of the application.

Several technological advancements have played a critical role in enabling intelligent product ecosystems. The adoption of RFID technologies, ubiquitous GPS, and order tracking software has significantly improved real-time asset visibility and traceability \citep{McFarlane2012IntelligentSupplyChain}. Additionally, the development of web services and cloud computing architectures has enhanced the scalability of networked intelligent products, allowing them to access and process large volumes of distributed data. These innovations have been instrumental in transforming traditional supply chain management, industrial automation, and IoT-driven infrastructure, enabling a shift from reactive systems to proactive, self-optimising networks.

A key consideration in the design of intelligent products is their ability to function within heterogeneous environments, where different product types, communication protocols, and decision models must interact seamlessly. International standards such as ISO/IEC 19944 (Data Flow for Connected Devices) \citep{ISO19944} and ISO 23247 (Digital Twin Frameworks for Manufacturing) \citep{ISO23247}is  provide a structured approach for enabling cross-domain interoperability in intelligent product ecosystems. By integrating these standards with multi-agent coordination and distributed intelligence models, intelligent products can operate with greater autonomy and efficiency, bridging the gap between physical assets and digital decision-making infrastructures.

While traditional IoT systems rely on cloud-based processing for decision-making, emerging distributed computing models are enabling products to operate with localised intelligence. Edge computing architectures allow intelligent products to process data and make decisions on-device, reducing latency, bandwidth dependence, and reliance on centralised control systems. This decentralised approach is particularly beneficial in applications that require low-latency responses, such as real-time process automation in smart factories and autonomous robotic systems in logistics and healthcare industries.

\subsection{Limitations of Existing Intelligent Product Architectures}

Despite notable advancements in the development of intelligent product systems \citep{Meyer2009Survey}, current architectures remain constrained by several fundamental limitations that restrict their ability to function autonomously, adapt dynamically, and operate across heterogeneous environments. The predominant focus of existing research and implementation has been on supply chain and industrial automation applications, where intelligent products primarily serve as data carriers and tracking mechanisms rather than as autonomous, decision-making entities. However, as intelligent products expand into broader domains, such as human-robot interaction, smart infrastructure, and consumer applications, their existing limitations become increasingly evident.

\subsubsection{Limitations in Intelligence and Autonomy}

Current intelligent product architectures predominantly operate at Level 1 Intelligence, where products can identify themselves, collect contextual data, and transmit information to centralised systems for processing. However, they lack the capability for higher-order reasoning and autonomous decision-making necessary for Level 2 Intelligence. While RFID-enabled products can track environmental conditions, trigger alerts, and relay data to cloud-based platforms, they are incapable of resolving disruptions independently or engaging in collaborative problem-solving with other intelligent products.

For intelligent products to be truly intelligent, they must transition from passive data collection and rule-based automation to collaborative but autonomous decision-making with real-time adaptation. Prior research has proposed product-driven supply threads, in which intelligent products autonomously negotiate transportation routes, optimise delivery costs, and arrange logistics with minimal human intervention \citep{Brintrup2011Aircraft}. However, practical implementation remains limited due to the absence of decentralised coordination frameworks and autonomous negotiation mechanisms, which restrict the ability of intelligent products to function as self-governing entities.

\subsubsection{Challenges in Interoperability and Plug-and-Play Capabilities}

Interoperability remains a critical challenge in the development of intelligent product ecosystems \citep{Valckenaers2009AgereVesere}. Many existing implementations operate within proprietary, manufacturer-specific silos, leading to fragmentation that prevents seamless data exchange between platforms \citep{McFarlane2012IntelligentSupplyChain}. The absence of universally accepted plug-and-play capabilities necessitates the use of custom integrations and structured API communication protocols, which significantly limits the scalability and flexibility of intelligent product networks.

Existing IoT architectures require structured, deterministic communication protocols, which contrast with the adaptive, unstructured nature of human communication and decision-making. While traditional IoT systems rely on fixed API specifications for interoperability, future intelligent product systems must develop flexible, self-organising communication frameworks that enable adaptive interactions without rigid standardisation. Achieving interoperability without strict standardisation remains a significant research challenge, particularly given the reluctance of large technology companies to collaborate on open, interoperable standards.

\subsubsection{Dependence on Centralised Architectures and Cloud-Based Decision-Making}

Existing intelligent product frameworks rely on centralised data architectures, where product data is transmitted to cloud servers or central control systems for aggregation, processing, and decision-making \citep{Meyer2011Production}. While this approach enables oversight and large-scale coordination, it introduces limitations in scalability, autonomy, and responsiveness. Centralised systems create bottlenecks, as all data and control flows must pass through a central point, adding communication and coordination delays. This dependence on central control reduces the ability of intelligent products to respond quickly to local events or operate independently in dynamic or disconnected environments.

Multi-agent systems (MAS) have been proposed as a means of enabling distributed intelligence and decentralised decision-making. However, current MAS implementations still depend on centralised orchestration, where autonomous agents operate within predefined system constraints and rely on central hubs for coordination \citep{McFarlane2002IPMC}. This dependency on centralised architectures limits the ability of intelligent products to function independently in disconnected environments or dynamically adapt to changing operational conditions.

For intelligent products to function beyond supply chain automation, they must transition toward self-sovereign architectures, wherein decision-making and operational intelligence are embedded at the object level \citep{Meyer2009Survey}. This shift requires products to process real-time data locally, execute autonomous decision-making without cloud dependence, and coordinate actions dynamically with other intelligent products. However, existing architectures lack the infrastructure to support fully decentralised intelligence, making them reliant on structured cloud-based computing models and centralised coordination frameworks.

\subsubsection{Limitations in Knowledge Sharing and Learning}

Another fundamental limitation of existing intelligent product architectures is their inability to support distributed knowledge sharing and collective learning mechanisms. While intelligent products are capable of collecting and transmitting data, they currently operate within isolated knowledge ecosystems, where learning is limited to predefined data inputs and static rules rather than continuous, collaborative knowledge exchange.

Intelligent products should be able to share learned experiences, collaboratively refine decision models, and form a collective intelligence network that enhances overall system efficiency \citep{Kiritsis2011PLM}. However, existing architectures do not support such collaborative, knowledge-sharing mechanisms, as they rely on static data models rather than self-evolving intelligence frameworks. This limitation is particularly pronounced in environments where intelligent products must interact with AI-driven agents, robots, and autonomous systems, requiring dynamic adaptation and real-time learning capabilities.

\subsubsection{Safety, Control, and Explainability in Intelligent Product Systems}

As intelligent products become increasingly autonomous, ensuring safe and controlled interactions with physical environments is of paramount importance. Unlike traditional IoT systems, which follow predefined rule-based automation, future intelligent products will need to interpret unstructured data, adapt behaviour dynamically, and make real-time context-aware decisions. However, this introduces significant concerns regarding safety, accountability, and explainability in decision-making processes.

For example, while large language models (LLMs) and AI-driven decision systems have demonstrated impressive advancements in natural language processing and contextual reasoning, they also introduce risks related to hallucination, biased reasoning, and unpredictable behaviour. If integrated into physical AI and embodied intelligence systems, these unpredictable behaviours could lead to erroneous or hazardous interactions with physical objects and human users.

Former Google CEO, Eric Schmidt, has highlighted the necessity of implementing fail-safe mechanisms that allow human operators to ``pull the plug'' \citep{schmidt2024fortune} on autonomous decision-making systems when their behaviour becomes incomprehensible or potentially harmful. Ensuring that intelligent products are designed with robust fail-safe protocols, human oversight mechanisms, and explainability frameworks will be essential in mitigating risks associated with autonomous operation in complex environments.

\subsection{Towards a New Model for Enabling Intelligent Products}

Existing intelligent product architectures remain constrained by their reliance on centralised coordination, limited interoperability, and insufficient support for autonomous decision-making. Although considerable progress has been made, these core challenges may impede broader adoption and more advanced real-world applications. In order to address these limitations, the next section will investigate key design principles and technological requirements, laying the groundwork for overcoming the current limitations and enabling more robust, adaptable intelligent product systems.

\clearpage
\section{The Intelligent Product 3.0: Rehashing the Foundation}
\label{sec:3}

The transition to Intelligent Product 3.0 represents a fundamental shift in how physical objects operate within digital ecosystems and interact with autonomous agents. While earlier iterations relied on centralised cloud computing and rule-based automation, Intelligent Product 3.0 leverages Web3 - encompassing blockchain, decentralised infrastructures and generative artificial intelligence - to create self-sovereign, economically active entities. These products autonomously interact with their environments, serving as independent agents within trustless, distributed networks. In this section, we revisit the foundational concepts of the Intelligent Product and explore how Web3 technologies provide the opportunity to expand its capabilities and reshape its autonomous interactions within distributed environments.

\subsection{Scope of an Intelligent Product}

The definition of an intelligent product has been well established in existing literature, often referring to physical objects equipped with identification, sensing, and decision-making capabilities (McFarlane et al., 2003; Wong et al., 2002). While the concept of intelligence within a product is defined, the term ``product'' itself requires further clarification.

A ``product'' is traditionally defined as a manufactured good - an item produced at scale with a consistent design and intended for commercial distribution, typically identified through global standards such as GS1 barcodes or Electronic Product Codes (EPCs). However, a product can also encompass custom-made, one-off items that lack a standard identifier. The foundational characteristic of an Intelligent Product is its ability to possess a unique identity; thus, for any product to be considered intelligent, it must first have a globally unique identifier, whether structured through established standards or assigned through alternative means.

With the advancement of IoT technologies, manufactured products increasingly incorporate onboard processing, enabling real-time data collection and connectivity. While many IoT devices meet some criteria of an intelligent product, they often rely on a centralised, rule-based architecture, limiting their ability to autonomously interact with their environment or participate in decentralised decision-making. IoT systems primarily function within closed, centrally controlled ecosystems, where communication is structured and directed only towards central applications, restricting their ability to interact dynamically with their environment or participate in decision-making about their own operation. This reliance on predefined rules and structured API frameworks prevents IoT devices from evolving into fully autonomous, intelligent products capable of self-directed adaptation and real-time collaboration.

Recent advancements in robotics, particularly humanoid robots, introduce a new level of product capability, extending beyond onboard processing to incorporate actuation and autonomous interaction. While humanoid robots are designed to exhibit greater autonomy, they are likely to inherit the centralised, rule-based characteristics of IoT, albeit with increasingly sophisticated capabilities driven by large language models (LLMs).

\begin{table}[ht]
\centering
\renewcommand{\arraystretch}{1.3} 
\setlength{\tabcolsep}{6pt} 
\begin{tabular}{|l|l|l|}
\hline
 & \textbf{No Actuation} & \textbf{Capable of Actuation} \\
\hline
\textbf{Cloud Processing Only} & Cloud-Connected Products & Cloud-Dependent Actuating Products \\
 & (e.g.\ cloud-based warranty registration, & (e.g.\ Remote-controlled robotic arms, \\
 & RFID/barcode for supply chain tracking, & smart home devices like thermostats) \\
 & barcode-based product lookup) & \\
\hline
\textbf{Hybrid Processing} & Smart IoT Products & Autonomous Products \& Robots \\
\textbf{(Onboard + Cloud)} & (e.g.\ AI-powered security cameras, & (e.g.\ Self-navigating drones, \\
 & environmental sensors, smart wearables) & industrial robots, humanoid robots) \\
\hline
\end{tabular}
\caption{Categories of Intelligent Products}
\label{tab:ipcategories}
\end{table}
In order to provide a clearer definition of the scope and evolution of intelligent products, we categorise Intelligent Products along two key dimensions as shown in Table~\ref{tab:ipcategories}. The categorisation of intelligent products in this framework considers two key dimensions: processing architecture and actuation capability. Processing architecture refers to where computational tasks are performed - either entirely in the cloud or distributed between onboard (local) processing and the cloud (hybrid processing). Products relying solely on cloud processing depend on external infrastructure for data analysis and decision-making, whereas hybrid processing products combine local computational capacity with cloud-based resources, enabling more autonomous and context-aware operations. Platforms such as itemit.com \citep{itemit2025} already incorporate autonomous RFID agents with hybrid processing architectures to manage local reader operations while synchronising data to the cloud, demonstrating the practical implementation of distributed intelligence at the edge. 

Actuation capability describes whether a product is limited to sensing and information provision or whether it is also equipped to perform physical actions on its environment. Products without actuation capability are informational in nature, while those with actuation capability can interact with and alter their surroundings through embedded mechanisms such as motors, actuators, or manipulators. This categorisation provides a framework for distinguishing between different types of intelligent products based on their functional roles and technical architectures. 

A separate category for products relying solely on onboard processing without cloud connectivity is intentionally excluded. Although such products exist, particularly in contexts requiring strict isolation for security or safety reasons, they are becoming increasingly uncommon in most commercial and industrial settings. This trend reflects growing expectations for continuous data exchange, integration with broader digital ecosystems, and the need for remote updates and monitoring. Products with purely local processing are often constrained in scalability, interoperability, and adaptability, rendering them less suited to the evolving landscape of intelligent products, which prioritises connectivity, integration, and dynamic collaboration.

\subsection{Key Characteristics of an Intelligent Product}

In order to overcome the limitations outlined in Section~2.3, intelligent product architectures must exhibit the following characteristics:

\paragraph{Definition 3: Characteristics of an Intelligent Product}
\begin{itemize}
    \item \textbf{Autonomy} -- The capacity to make independent decisions and dynamically adapt to new conditions without reliance on centralised control.
    \item \textbf{Consensus} -- The ability to negotiate and coordinate actions across multi-stakeholder supply chains, reducing reliance on predefined control structures.
    \item \textbf{Interoperability} -- The ability to seamlessly communicate across different ecosystems, regardless of structured API limitations, proprietary constraints, or platform dependencies.
    \item \textbf{Intelligence Sharing} -- Moving beyond static, rules-based intelligence to adaptive, self-improving knowledge systems, where intelligent products can collaborate, exchange insights, and refine their capabilities dynamically.
    \item \textbf{Safety and Control} -- The implementation of fail-safe mechanisms, real-time monitoring frameworks, and explainability models to ensure that autonomous products behave predictably and securely.
\end{itemize}

Recent advancements in Web3 technologies \citep{Wood2014Ethereum}, including blockchain and decentralised protocols, have significantly addressed most of the characteristics above. By leveraging trustless and distributed infrastructures \citep{Wang2022Web3}, these technologies reduce reliance on centralised control and proprietary constraints. This decentralisation enables intelligent products to operate with greater autonomy, improved interoperability, and enhanced security. While an intelligent product need not exhibit every one of these characteristics to be considered as such, incorporating each additional characteristic further enhances the product's adaptive capacity, decision-making ability, and overall effectiveness.

For intelligent products to fully realise their potential across manufacturing, logistics, human-machine interaction, and consumer applications, they must move beyond static, rule-based automation and incorporate self-learning and decentralised intelligence frameworks. The ability to negotiate, coordinate, and dynamically adapt to evolving conditions will be critical for creating scalable and resilient intelligent product ecosystems. Even with these capabilities, overcoming current challenges in autonomy, interoperability, and intelligence sharing remains essential for developing next-generation intelligent products capable of real-time decision-making, adaptive collaboration, and seamless integration into complex, multi-agent environments.

\subsection{Deploying Intelligent Products}

In this section, the evolution of intelligent product architecture is examined in response to technological developments. Intelligent Product 1.0 was based on unique identifiers, particularly EPC, centralised databases, and rule-based decision agents. This model enabled products to sense and share basic data but was limited by server-based processing and rigid communication structures. Its reliance on centralised systems constrained autonomy, adaptability, and scalability.

Intelligent Product 2.0 introduced advances such as Universally Unique Identifiers (UUID), digital twins, cloud-based storage, and AI-augmented digital twins. These developments expanded the ability of the product to represent and communicate its state digitally, while enabling remote monitoring and data analysis through cloud platforms. However, Intelligent Product 2.0 remained dependent on centralised architectures, facing challenges with interoperability, fragmented standards, and limited infrastructure for large-scale deployment.

Earlier versions of intelligent product architectures were therefore not easily deployable at scale; the challenges were not only theoretical but also practical and technological, including issues of infrastructure, interoperability, and system readiness. In contrast, recent advances in decentralised technologies, artificial intelligence, and interoperable digital ecosystems have addressed many of these barriers. The underlying technical and infrastructural conditions are now in place, making Intelligent Product 3.0 both feasible and timely as shown Table~\ref{tab:ipevolution}.
\clearpage

\begin{table}[ht]
\centering
\renewcommand{\arraystretch}{1.3} 
\setlength{\tabcolsep}{6pt} 
\begin{tabular}{|p{4cm}|p{3.5cm}|p{3.5cm}|p{3.5cm}|}
\hline
\textbf{} & \textbf{Intelligent Product 1.0} & \textbf{Intelligent Product 2.0} & \textbf{Intelligent Product 3.0} \\
\hline
\textbf{1. Possesses a unique identity} & EPC & EPC, UUID & Decentralised ID (DID) \\
\hline
\textbf{2. Capable of communicating effectively with its environment} & PML & EPCIS and Digital Twin & Extended DPP \\
\hline
\textbf{3. Can retain or store data about itself} & Server-based & Cloud-based & Blockchain-based, DePIN \\
\hline
\textbf{4. Deploys a language to display its features, etc.} & Predefined, structured communication & AI-Augmented Digital Twin (structured) & Agentic AI with unstructured, adaptive communication \\
\hline
\textbf{5. Capable of participating in or making decisions relevant to its own destiny} & Rule-based agents & Fine-tuned AI models & Autonomous intelligent agents (LLM \& decentralised AI) \\
\hline
\end{tabular}
\caption{Evolution of Intelligent Product Architectures}
\label{tab:ipevolution}
\end{table}

Intelligent Product 3.0 integrates blockchain, decentralised networks, digital product passports, and AI agents to enhance autonomy, consensus, interoperability, intelligence sharing, and safety. Blockchain ensures decentralised trust, data integrity, and transparency, addressing security and interoperability challenges \citep{Nakamoto2008Bitcoin,buterin2013ethereum, Wood2014Ethereum}. By removing reliance on centralised intermediaries, blockchain allows intelligent products to interact directly, supporting self-executing smart contracts that enforce agreements without human intervention.

Digital Product Passports (DPPs) and Digital Link Standard \citep{GS1DigitalLink2018}, linked to Decentralised Identifiers (DIDs) \citep{W3C2022DIDcore}, provide immutable records for product history, maintenance, and compliance \citep{Jansen2022DPPoverview}. This ensures products retain verifiable identities throughout their lifecycle, reducing fraud and enhancing traceability \citep{Saberi2019BlockchainSSC}. With these digital passports information, manufacturers, consumers, and regulators can access verifiable records, improving accountability and sustainability \citep{EC2022DPP}.

Decentralised Physical Infrastructure Networks (DePIN) create distributed IoT backbones, reducing dependency on centralised providers \citep{Ballandies2023DePIN}. These networks leverage community-driven infrastructure, where contributors provide resources such as IoT connectivity and edge computing in exchange for blockchain-based incentives \citep{Christidis2016BlockchainIoT}. DePINs allow intelligent products to remain operational within decentralised environments, enabling real-time intelligence sharing and reducing single points of failure.

Agentic AI, supported by large language models (LLMs) and reinforcement learning, enables products to autonomously assess data, negotiate interactions, and execute transactions \citep{Nguyen2024AutonomousAIBlockchain}. By embedding AI within intelligent products, decision-making is dynamic and context-aware, reducing reliance on pre-programmed rules. AI-driven products can negotiate with other agents, optimise resource allocation, and respond to changing conditions in real-time, enhancing adaptability \citep{RussellNorvig2021AI}.

Deploying Intelligent Product 3.0 requires integrating these technologies to enable self-sufficient, secure, and scalable product ecosystems. Decentralisation ensures trust, peer-to-peer communication fosters adaptability, and autonomous AI enhances decision-making. This transition represents a paradigm shift from cloud-reliant, centrally controlled products to networked intelligent agents that can operate independently while maintaining interoperability and security.

\subsection{New Capabilities Enabled by Intelligent Product 3.0}

The Intelligent Product 3.0 paradigm introduces several transformative opportunities, leveraging decentralised infrastructure, AI integration, and tokenisation to create autonomous, economically active systems. These opportunities redefine how intelligent products interact, transact, and collaborate across digital and physical environments.

\paragraph{1. Tokenised Knowledge and AI-Driven Economies}
Intelligent Product 3.0 enables products to act as economic agents that can autonomously exchange knowledge, insights, and services. Intelligent products transition from passive assets to economic agents by leveraging tokenised incentives. Through decentralised AI marketplaces, products can trade predictive analytics, operational data, and optimisation insights directly with other products or AI agents \citep{Goertzel2017SingularityNET}. Smart contracts enable micropayments for data-sharing, allowing intelligent products to communicate with AI agents without human intervention. This model facilitates real-time monetisation of machine intelligence, promoting seamless interaction between diverse systems. Tokenised payments allow intelligent products to function across multiple ecosystems, enabling frictionless interoperability across industries.

\paragraph{2. Global Mobility and Interoperability of Products}
Intelligent Product 3.0 allows products to move seamlessly across different platforms, organisations, and regions while retaining their identity, access rights, and operational data. By using decentralised identifiers (DIDs), Verifiable Credentials (VCs), and extended Digital Product Passports (DPPs), products can authenticate themselves, prove their status, and securely exchange information without relying on centralised databases. This capability enables an intelligent product to be recognised, trusted, and functional regardless of location or system boundaries - supporting scenarios such as cross-border logistics, multi-owner supply chains, and distributed operations. Zero-knowledge proofs (ZKPs) protect sensitive information during authentication, while decentralised storage ensures that a product's data is resilient and accessible even without a central authority.

\paragraph{3. Collaboration with AI Agents, Embodied Robots, and Machines}
Intelligent Product 3.0 enables products to directly collaborate with autonomous robots, embodied AI, and other intelligent machines, extending their role from data providers to participants in physical action. By integrating decentralised identity (DID), smart contracts, and Decentralised Physical Infrastructure Networks (DePIN), products can negotiate tasks, schedule operations, and coordinate with embodied AI systems such as drones, robotic arms, or industrial robots. This capability removes the need for centralised intermediaries to coordinate actions, allowing intelligent products to initiate and oversee tasks in the real world. It supports distributed automation, reduces reliance on central control, and enables decentralised governance across supply chains and industrial ecosystems.

\paragraph{}The following Table~\ref{tab:opportunities-usecases} provides example use cases illustrating how Intelligent Product 3.0 capabilities align with the opportunities outlined above. These examples demonstrate practical applications, highlighting autonomous interactions, economic participation, and collaboration across various ecosystems.

\begin{table}[ht]
\centering
\renewcommand{\arraystretch}{1.3} 
\setlength{\tabcolsep}{6pt} 
\begin{tabular}{|p{3cm}|p{12cm}|} 
\hline
\textbf{Opportunity} & \textbf{Use Cases} \\
\hline
\textbf{Tokenised Knowledge and AI-Driven Economies} & 
\vspace{-\baselineskip} 
\begin{itemize}
    \item Instant Tool Use and Authentication: Robots and smart machines autonomously purchase verified manuals via tokenised micropayments and confirm genuine tools via zero-knowledge proofs.
    \item Automated Repair at Point of Usage: Robots and smart machines pay manufacturers for certified repair instructions, record each repair, and monetise repair data.
    \item Monetisable Innovation Sharing: Intelligent products share new assembly or efficiency methods on decentralised marketplaces, earning tokens from other AI agents.
\end{itemize} \\
\hline
\textbf{Global Mobility and Interoperability of Products} & 
\vspace{-\baselineskip} 
\begin{itemize}
    \item Immutable Product Passports: Decentralised ledgers store tamper-proof product histories, removing the need for a single authority and enabling reliable authenticity checks.
    \item Autonomous Validation for High-Value Goods: Everyday items and premium products (e.g., wine) verify provenance through immutable records, thwarting counterfeits without central oversight.
    \item Smart Recycling and Disposal: Robots use decentralised, immutable disposal instructions, ensuring consistent compliance across ecosystems and reducing reliance on centralised control.
\end{itemize} \\
\hline
\textbf{Collaboration with AI Agents, Embodied Robots, and Machines} & 
\vspace{-\baselineskip} 
\begin{itemize}
    \item Collaborative Embodied AIs: Intelligent household devices autonomously communicate and coordinate tasks with robots from different manufacturers (e.g., Tesla, Dyson, Samsung) using decentralised AI protocols, optimising chores and resource usage.
    \item Social Network for Embodied AIs: Robots, machines and devices exchange experiences and ratings on repair instructions via decentralised channels, collectively enhancing repair quality.
    \item Smart Recycling and Disposal: Robots, machines and devices autonomously manage their own recycling processes using environmental data, ensuring sustainable disposal. Intelligent items instruct robots on proper handling via standardised environmental impact guidelines.
\end{itemize} \\
\hline
\end{tabular}
\caption{Opportunities and Use Cases for Intelligent Products}
\label{tab:opportunities-usecases}
\end{table}

\clearpage
\section*{Conclusions}

Intelligent Product 3.0 represents a shift toward autonomous, economically active, and self-sustaining systems that move beyond traditional, centrally controlled architectures. By integrating decentralised AI, blockchain authentication, and tokenised economic models, products no longer act as passive assets but function as intelligent, decision-making entities within distributed machine economies. This transition enables self-sovereign operation, enhanced interoperability, and ongoing collaboration between AI agents, embodied AI, and physical products.

Three key conditions now make Intelligent Product 3.0 feasible. First, the technological infrastructure - including decentralised identifiers, blockchain systems, AI agents, and interoperable data frameworks - is already in place. Second, there is growing demand across industries for autonomous, self-governing products capable of operating in decentralised environments and participating in economic activities. Third, initiatives such as umin.ai are emerging to integrate these technological capabilities and application needs, bringing together enabling tools, platforms, and practical use cases.

Future efforts will need to address challenges related to regulatory compliance, governance, and system reliability to ensure secure, verifiable, and trustworthy implementations. If these challenges are tackled, Intelligent Product 3.0 can establish a scalable, decentralised ecosystem that advances beyond passive tracking and centralised oversight, moving toward a truly self-sovereign, AI-first environment.

Future research will explore the implementation of this conceptual framework by developing an intelligence marketplace (including DePIN), a public library of intelligent products, and AI-to-AI communication methods to support self-sustaining collaboration, as part of the umin.ai Foundation \citep{uminai2025} initiative advancing decentralised AI agents and Web3 intelligence standards.

\bibliographystyle{apalike}      
\bibliography{bibliography}      
\end{document}